\documentclass[%
 reprint,
 amsmath,amssymb,
 aps,
]{revtex4-1}

\usepackage[dvipdfmx]{graphicx}% Include figure files
\usepackage{dcolumn}% Align table columns on decimal point
\usepackage{bm}% bold math
\usepackage{braket}

\usepackage{ulem}
\usepackage{color}

\newcommand{\muetoee}{$\mu^-e^-\rightarrow e^-e^-$~}

\begin{document}
\preprint{}
%\linenumbers

\title{
Improved analyses for $\mu^-e^-\rightarrow e^-e^-$ in muonic atoms by contact interactions}% Force line breaks with \\
%\thanks{}%

\author{Yuichi Uesaka$^1$, Yoshitaka Kuno$^1$, Joe Sato$^2$, Toru Sato$^{1,3}$, and Masato Yamanaka$^4$}
\affiliation{$^1$Department of Physics, Osaka University, Toyonaka, Osaka 560-0043, Japan\\
$^2$Physics Department, Saitama University, 255 Shimo-Okubo, Sakura-ku, Saitama, Saitama 338-8570, Japan\\
$^3$J-PARC Branch, KEK Theory Center, Institute of Particle and Nuclear Studies, KEK, Tokai, Ibaraki, 319-1106, Japan\\
$^4$Department of Physics, Nagoya University, Nagoya 464-8602, Japan}
% \altaffiliation[Also at ]{Department of Physics, Osaka University.}%Lines break automatically or can be forced with \\
%\affiliation{%

%\collaboration{}%\noaffiliation

%\author{}
% \homepage{}
%\affiliation{
%\affiliation{
%\affiliation{%


\date{\today}% It is always \today, today,

\begin{abstract}
The charged lepton flavor violating (CLFV) processes of $\mu^-e^-\rightarrow e^-e^-$ decay 
by four Fermi contact interactions in a muonic atom for various atoms are investigated.
%for the case 
% YK modified the following 20160226
The wave functions of bound and scattering state leptons are properly 
treated by solving Dirac equations with Coulomb interaction of the finite nuclear charge distributions.
This new effect contributes significantly in particular for heavier atoms, 
where the obtained decay rate is about one order of magnitude larger than 
the previous estimation for $^{208}$Pb.
We find that, as the atomic number $Z$ increases,
the $\mu^-e^-\rightarrow e^-e^-$ decay rates increase more rapidly %than $Z^3$,
 than the result of the previous work of $Z^3$,
suggesting this decay as one of the promising processes to search for CLFV  interaction.
\end{abstract}

\maketitle

\onecolumngrid

%\tableofcontents

\section{Introduction}

The charged lepton flavor violating (CLFV) processes are known 
to provide 
%an % commented out by YK 20160226
important signals on  physics beyond the Standard Model (SM).
% YK modified the following 201502026
The analysis of search for $\mu^+\rightarrow e^+\gamma$ decays in the cosmic-ray muons by Hincks and Pontecorvo 
in 1947 \cite{Hincks1948} has given the first upper limit on the branching ratio of CLFV processes.
Since then, the upper limits of the branching ratios of CLFV processes have been improved and now reach around the orders 
of $10^{-12} \sim 10^{-13}$ \cite{Kuno2001,Kuno2013}.
These upper limits put stringent constraints on various theoretical 
models beyond the SM.
These CLFV searches include the processes
such as $\mu^+\rightarrow e^+\gamma$, $\mu^+\rightarrow e^+e^-e^+$ and $\mu^- N \rightarrow e^- N$ conversion in a muonic atom. Recently, another process of $\mu^-e^-\rightarrow e^-e^-$ decay in a muonic atom has been proposed by 
Koike {\it et al.} in 2010 \cite{Koike2010}. 
A unique feature of this process is that both photonic and contact leptonic interaction can be proved,
 and an experimentally clean signal is expected because the sum of the energies of two electrons
 is restricted to the muon mass minus the binding energy of the muon in a muonic atom. 
The measurement of this process is planned
in the COMET Phase-I experiment in J-PARC \cite{KEKrep}.

In Ref.~\cite{Koike2010}, the decay rate of muonic atom was evaluated by using the non-relativistic 
bound state wave functions of muon and electron and the plane wave approximation of the final electrons.
It was shown the decay rate increases with the atomic number $Z$ as $\Gamma \sim Z^3$.
Therefore heavy muonic atoms %with the enhancement of decay rate
would provide a great opportunity of CLFV search.
However, as is well known, the effects of the Coulomb interaction is significant for the ordinary decay of bound muons in heavy nuclei \cite{Huff1961,Watanabe1987}.
Since the quantitative evaluation of the decay process is needed in order to disentangle the mechanism of CLFV interaction, it is important to update the estimations of Ref.~\cite{Koike2010} by taking into account the effects of the Coulomb interactions for the relativistic leptons.
The importance of the Coulomb distortion for the $\mu^--e^-$ conversion process in
a muonic atom has been reported in Refs.~\cite{Kitano2002,Shanker1979,Czarnecki1998}.
For $\mu^--e^-$ conversion process where the nucleus stays intact,
it is sufficient to consider the s-wave muon and  electron states.
For $\mu^-e^-\rightarrow e^-e^-$ decay of muonic atom,
on the other hand, two electrons with the energy of approximately a half of muon mass are emitted 
in the final state. 
The angular momentum of the each electrons is not limited in this process.
A formalism of the $\mu^-e^-\rightarrow e^-e^-$ decay with the partial wave expansion of leptons is necessary,
 as has been common in the nuclear beta decay and muon capture reactions \cite{koshigiri1979}.

In Section \ref{sec:Formalism}, we summarize the relevant effective Lagrangian for 
the $\mu^-e^-\rightarrow e^-e^-$ process and develop a formula of the decay rate using the 
partial wave expansion of the lepton wave function.
Our refined estimations of the $\mu^-e^-\rightarrow e^-e^-$ decay rate for the muonic atom is presented in Section \ref{sec:Results}.
Finally our conclusion is given in Section \ref{sec:Conclusion}.

\section{Formulation \label{sec:Formalism}}
The effective Lagrangian of the CLFV process $\mu^-e^-\rightarrow e^-e^-$ is given as,
\begin{align}
\mathcal{L}_I=&\mathcal{L}_\mathrm{photo}+\mathcal{L}_\mathrm{contact}, \\
\mathcal{L}_\mathrm{photo}=&-\frac{4G_F}{\sqrt{2}}m_\mu
\left[A_R\overline{e_L}\sigma^{\mu\nu}\mu_R+A_L\overline{e_R}\sigma^{\mu\nu}\mu_L\right]F_{\mu\nu}+[h.c.], \\
\mathcal{L}_\mathrm{contact}=&-\frac{4G_F}{\sqrt{2}}
[g_1(\overline{e_L}\mu_R)(\overline{e_L}e_R)
+g_2(\overline{e_R}\mu_L)(\overline{e_R}e_L) \nonumber\\
&+g_3(\overline{e_R}\gamma_\mu\mu_R)(\overline{e_R}\gamma^\mu e_R)
+g_4(\overline{e_L}\gamma_\mu\mu_L)(\overline{e_L}\gamma^\mu e_L) \nonumber\\
&+g_5(\overline{e_R}\gamma_\mu\mu_R)(\overline{e_L}\gamma^\mu e_L)
+g_6(\overline{e_L}\gamma_\mu\mu_L)(\overline{e_R}\gamma^\mu e_R)]+[h.c.],
\label{eq:shortrange}
\end{align}
where $G_F=1.166\times10^{-5}$GeV$^{-2}$ is the Fermi coupling constant, 
and $A_{R,L}$ and $g_i$'s $(i=1,2,\cdots,6)$ are dimensionless coupling constants.
The left and right handed fields $\psi_{L/R}$ are given as $\psi_{L/R}=P_{L/R}\psi$ 
with $P_{L/R}=(1\mp \gamma_5)/2$.
The effective Lagrangian consists of two parts.
The first part, $\mathcal{L}_\mathrm{photo}$, represents the photonic interaction of $\mu\rightarrow e\gamma$ types, which 
generates the long range $\mu-e$ interaction with one photon exchange between a muon and an electron.
The second part, $\mathcal{L}_\mathrm{contact}$, is the four Fermi
 interaction.
In this work, we concentrate on the contact interaction as our first attempt to examine the role of 
Coulomb interaction  on the \muetoee decay of muonic atoms.

We evaluate the decay rate of two-electron emission of the muonic 
atom within the independent particle picture of the muonic atom and the final state.
The transition amplitude is given by the matrix element of the effective CLFV interaction 
in Eq. (\ref{eq:shortrange})
\begin{eqnarray}
M(\bm{p}_1,s_1,\bm{p}_2,s_2 ;\alpha_\mu,s_\mu,\alpha_e,s_e) & \equiv&
\int d^3r\braket{e^{s_1}_{\bm{p}_1}e^{s_2}_{\bm{p}_2}|\mathcal{L}_\mathrm{contact}|\mu^{s_\mu}_{\alpha_\mu}e^{s_e}_{\alpha_e}} \nonumber\\
& = &
-\frac{4G_F}{\sqrt{2}}\sum_{i=1}^{6}g_i\left[\int d^3r
\overline{\psi}^{e(-)}_{\bm{p}_1,s_1}(\bm{r})O_i^A\psi_{\alpha_\mu,s_\mu}^\mu(\bm{r})
\overline{\psi}^{e(-)}_{\bm{p}_2,s_2}(\bm{r})O_i^B\psi_{\alpha_e,s_e}^e(\bm{r})-(1\leftrightarrow 2)\right],
\label{eq:MatrixElement}
\end{eqnarray}
where $\psi^{e(-)}_{\bm{p},s}(\bm{r})$ is the wave function of a scattering electron with its momentum $\bm{p}$ and spin $s$.
The superscript $(-)$ represents the incoming wave boundary condition.
The wave functions of bound leptons are denoted as $\psi^l_{\alpha,s}$ with $l=\mu,e$, 
spin $s$ and $\alpha=n,\kappa$.
Here, $\kappa$ represents both the orbital and the total angular momentum simultaneously \cite{rose1961,rose1995elementary}.
The second term $(1\leftrightarrow 2)$ in Eq. (\ref{eq:MatrixElement}) is the exchange 
term obtained from the first term by exchanging the quantum numbers of final electrons.
The Dirac matrix $O_i^A$ and $O_i^B$ for each $g_i$ in Eq. (\ref{eq:shortrange}) are given as
\begin{align}
O_1^A=O_1^B=P_R, &\, O_2^A=O_2^B=P_L, \nonumber\\
O_3^A=O_5^A=\gamma_\mu P_R, &\, O_4^A=O_6^A=\gamma_\mu P_L, \nonumber\\
O_3^B=O_6^B=\gamma^\mu P_R, &\, O_4^B=O_5^B=\gamma^\mu P_L.
\end{align}

We assume that the muon bound state is in the $n=1,\kappa=-1$ state
denoted simply by $\alpha_\mu=1S$.
Since orbit of the bound muon is about 200 times smaller than that of electron, 
the $n=1,\kappa=-1$  electron bound state gives the main contribution to the decay 
rate of the muonic atom as long as we consider the contact interaction in Eq. (\ref{eq:shortrange}). 
The  \muetoee decay rate of a muonic atom is given,
with possible contributions of electron bound states of $\kappa_e=-1$ and any $n$ included, 
as follows:
\begin{eqnarray}
\Gamma &=&
\frac{1}{2}\left(\sum_{s_1,s_2}\int\frac{d^3p_1d^3p_2}{(2\pi)^32E_1(2\pi)^32E_2}\right)
\left(\frac{1}{2}\sum_{s_\mu,s_e,n}\right) \nonumber\\
&\times& 2\pi\delta(E_{p_1}+E_{p_2}-m_\mu-m_e+B_\mu+B_e^n) 
%\nonumber\\
%\times 
\left|M(\bm{p}_1,s_1,\bm{p}_2,s_2;1S,s_\mu,nS,s_e)\right|^2, \label{eq:decayrate}
\end{eqnarray}
where $B_\mu$ and $B^n_e$ are the binding energies of the muon and electron in a muonic atom and 
$E_{p_i}$ is an energy of one of the electrons with its momentum $p_i$.
Here the initial muon spins are averaged.
The normalization of the bound state wave function is given as
\begin{eqnarray}
\int d^3\bm{r}{\psi^l_{\alpha,s}}^\dagger(\bm{r})\psi^{l}_{\alpha,s'}(\bm{r})
&= &\delta_{\alpha,\alpha'}\delta_{s,s'},
\end{eqnarray}
and the scattering wave function is normalized as
\begin{eqnarray}
\int d^3\bm{r}\psi^{e(-)\dagger}_{\bm{p},s}(\bm{r})\psi^{e(-)}_{\bm{p}',s'}(\bm{r})
& = & 2E_p(2\pi)^3\delta^3(\bm{p}-\bm{p}')\delta_{s,s'}.
\end{eqnarray}
The double differential decay rate with respect to the
electron energy and the angle $\theta$ between emitted electrons is given as
\begin{eqnarray}
\frac{d^2\Gamma_n}{dE_{p_1}d\cos\theta} & = &
\frac{4\pi\cdot 2\pi}{8(2\pi)^5}
|\bm{p}_1||\bm{p}_2|\sum_{s_1,s_2,s_\mu,s_e}
\left|M(\bm{p}_1,s_1,\bm{p}_2,s_2;1S,s_\mu,nS,s_e)\right|^2,
\label{eq:dgde}
\end{eqnarray}
where $E_{p_2}=-E_{p_1}+m_\mu+m_e-B_\mu-B_e^n$ and related to the total decay rate as
\begin{align}
\Gamma=\frac{1}{2}\sum_{n}\int_{m_e}^{E_\mathrm{max}^n}dE_{p_1}\int_{-1}^{1}d\cos\theta \frac{d^2\Gamma_n}{dE_{p_1}d\cos\theta},
\end{align}
where $E_\mathrm{max}^n=m_\mu - B_\mu - B_e^n$.
The transition matrix element $M$ is evaluated by using the
partial wave expansion of the electron scattering state.
The electron scattering state with the incoming boundary condition is expressed as
\begin{eqnarray}
\psi^{e(-)}_{\bm{p},s}(\bm{r})
&=&
\sum_{\kappa,\nu,m}4\pi i^{l_\kappa}(l_\kappa,m,1/2,s|j_\kappa,\nu)Y_{l_\kappa,m}^*(\hat{p})e^{-i\delta_\kappa}\psi^\kappa_{p,\nu}(\bm{r}),
\end{eqnarray}
where $\delta_\kappa$ is a phase shift for partial wave $\kappa$.
$\left(l_\kappa,m,1/2,s|j_\kappa,\nu\right)$ and $Y_{l_\kappa,m}(\hat{p})$ are Clebsch-Gordan coefficients and spherical harmonics, respectively.
Furthermore, the wave function $\psi^\kappa_{p,\nu}(\bm{r})$, where the subscripts $p$, $\nu$ mean a momentum 
of the electron and a spin of the partial wave, 
is written 
with the radial part $g_p^\kappa(r),f_p^\kappa(r)$ and the angular-spin 
part $\chi_\kappa$ \cite{rose1961,rose1995elementary} as follows,
\begin{align}
\psi^\kappa_{p,\nu}(\bm{r})=
\begin{pmatrix}
g_p^\kappa(r)\chi_\kappa^\nu(\hat{r}) \\
if_p^\kappa(r)\chi_{-\kappa}^\nu(\hat{r})
\end{pmatrix}.
\label{eq:scattering}
\end{align}
Similarly, the bound state wave function is given as
\begin{align}
\psi^l_{\alpha,s}(\bm{r})=
\begin{pmatrix}
g_{n,l}^\kappa(r)\chi_{\kappa}^s(\hat{r}) \\
if_{n,l}^\kappa(r)\chi_{-\kappa}^s(\hat{r})
\end{pmatrix},
\label{eq:bound}
\end{align}
where the subscript $l$ is for muon $l=\mu$ or electron $l=e$ and $s$ is a spin of the lepton.
The radial wave functions $g^\kappa(r)$ and $f^\kappa(r)$ are obtained by solving following 
Dirac equation with the Coulomb potential $V_{\rm{C}}(r)$ for appropriate boundary condition,
\begin{align}
\frac{dg^\kappa(r)}{dr}+\frac{1+\kappa}{r}g^\kappa(r)-\left(E+m+eV_{\rm{C}}(r)\right)f^\kappa(r)=& 0, \\
\frac{df^\kappa(r)}{dr}+\frac{1-\kappa}{r}f^\kappa(r)-\left(E-m+eV_{\rm{C}}(r)\right)g^\kappa(r)=& 0.
\label{eq:Dirac_eq}
\end{align}

Using the partial wave expansion of the scattering wave function
the transition amplitude can be written as follows
\begin{align}
M(\bm{p}_1,s_1,\bm{p}_2,s_2;1S,s_\mu,nS,s_e)
=& 2\sqrt{2}G_F
\sum_{\kappa_1,\kappa_2,\nu_1,\nu_2,m_1,m_2}
(4\pi)^2(-i)^{l_{\kappa_1}+l_{\kappa_2}}
e^{i\left(\delta_{\kappa_1}+\delta_{\kappa_2}\right)}
\nonumber\\
&\times Y_{l_{\kappa_1},m_1}(\hat{p}_1)Y_{l_{\kappa_2},m_2}(\hat{p}_2)
(l_{\kappa_1},m_1,1/2,s_1|j_{\kappa_1},\nu_1)(l_{\kappa_2},m_2,1/2,s_2|j_{\kappa_2},\nu_2) \nonumber\\
&\times\sum_{J,M}(j_{\kappa_1},\nu_1,j_{\kappa_2},\nu_2|J,M)(j_{-1},s_\mu,j_{-1},s_e|J,M) \nonumber\\
&\times\frac{2\sqrt{(2j_{\kappa_1}+1)(2j_{\kappa_2}+1)}}{4\pi}\sum_{i=1}^{6}
g_iW_i(J,\kappa_1,\kappa_2,E_{p_1}).
\end{align}
Here $W_i(J,\kappa_1,\kappa_2,E_{p_1})$ is the transition matrix element for the  $g_i$ term that 
includes both direct and exchange terms.
We introduce the function $Z_{ABCD}(L,S,J)$, which consists of the radial overlap integral, 
9j and parity Clebsch-Gordan coefficients as
\begin{align}
Z_{ABCD}(L,S,J)=& \int_{0}^{\infty}drr^2A_{p_1}^{\kappa_1}(r)B_{1,\mu}^{-1}(r)C_{p_2}^{\kappa_2}(r)D_{n,e}^{-1}(r) \nonumber\\
&\times\sqrt{\left(2l^A_{\kappa_1}+1\right)\left(2l^B_{-1}+1\right)\left(2l^C_{\kappa_2}+1\right)\left(2l^D_{-1}+1\right)} \nonumber\\
&\times\left(l^A_{\kappa_1},0,l^C_{\kappa_2},0|L,0\right)\left(l^B_{-1},0,l^D_{-1},0|L,0\right) \nonumber\\
&\times
\begin{Bmatrix}
l^A_{\kappa_1} & 1/2 & j_{\kappa_1} \\
l^C_{\kappa_2} & 1/2 & j_{\kappa_2} \\
L & S & J
\end{Bmatrix}
\begin{Bmatrix}
l^B_{-1} & 1/2 & 1/2 \\
l^D_{-1} & 1/2 & 1/2 \\
L & S & J
\end{Bmatrix}.
\end{align}
Here $A$ and $C$ represent the electron scattering states with momentum $p_1$ and $p_2$ and $B$ and $D$ represent the bound states of muon and electron.
The radial wave functions $A(r), B(r), C(r)$ and $D(r)$ are either $g(r)$ or $f(r)$ introduced in 
Eqs. (\ref{eq:scattering}) and (\ref{eq:bound}).
The angular momentum $l^h_{\kappa}$ is defined as
\begin{align}
l^h_\kappa=
\begin{cases}
l_{+\kappa} & \ \mbox{for} \ \ h=g, \\
l_{-\kappa} & \ \mbox{for} \ \ h=f.
\end{cases}
\end{align}

The amplitude $W_i$ for $i=1,\cdots,6$ are written by using linear combination of $Z$ as
\begin{align}
W_1(J)=&\frac{1}{2}\left\{X^-_\alpha(J,0,J)-X^+_\beta(J,0,J)+i\left[Y^+_\alpha(J,0,J)+Y^+_\beta(J,0,J)\right]\right\}, \\
W_2(J)=&\frac{1}{2}\left\{X^-_\alpha(J,0,J)-X^+_\beta(J,0,J)-i\left[Y^+_\alpha(J,0,J)+Y^+_\beta(J,0,J)\right]\right\}, \\
W_3(J)=& 2\left\{X^-_\alpha(J,0,J)+X^+_\beta(J,0,J)-i\left[Y^+_\alpha(J,0,J)-Y^+_\beta(J,0,J)\right]\right\}, \\
W_4(J)=& 2\left\{X^-_\alpha(J,0,J)+X^+_\beta(J,0,J)+i\left[Y^+_\alpha(J,0,J)-Y^+_\beta(J,0,J)\right]\right\}, \\
W_5(J)=& 3\sum_{L=|J-1|}^{J+1}X^-_\beta(L,1,J)-X^+_\alpha(J,0,J)+i\left[3\sum_{L=|J-1|}^{J+1}Y^-_\alpha(L,1,J)+Y^-_\beta(J,0,J)\right], \\
W_6(J)=& 3\sum_{L=|J-1|}^{J+1}X^-_\beta(L,1,J)-X^+_\alpha(J,0,J)-i\left[3\sum_{L=|J-1|}^{J+1}Y^-_\alpha(L,1,J)+Y^-_\beta(J,0,J)\right],
\end{align}
with
\begin{align}
X^\pm_\alpha(L,S,J)=& Z_{gggg}(L,S,J)+Z_{ffff}(L,S,J)\pm\left[Z_{gfgf}(L,S,J)+Z_{fgfg}(L,S,J)\right], \\
X^\pm_\beta(L,S,J)=& Z_{ggff}(L,S,J)+Z_{ffgg}(L,S,J)\pm\left[Z_{gffg}(L,S,J)+Z_{fggf}(L,S,J)\right], \\
Y^\pm_\alpha(L,S,J)=& Z_{ggfg}(L,S,J)-Z_{ffgf}(L,S,J)\pm\left[Z_{fggg}(L,S,J)-Z_{gfff}(L,S,J)\right], \\
Y^\pm_\beta(L,S,J)=& Z_{gggf}(L,S,J)-Z_{fffg}(L,S,J)\pm\left[Z_{gfgg}(L,S,J)-Z_{fgff}(L,S,J)\right].
\end{align}
Since we assume the bound states of muon and electron are both in the $\kappa=-1$ state, the total angular momentum $J$ can be $J=0$ or $1$ and $X_\mathrm{\alpha(\beta)}^\pm$ and $Y_{\alpha(\beta)}^\pm$ are non zero only for even $L$ and odd $L$ respectively.
It is noticed that only $S=0$ term contributes for $W_1$, $W_2$, $W_3$ and $W_4$, while both $S=0$ and $1$ terms contribute for $W_5$ and $W_6$.

After summing the spins of leptons, we yield the differential transition rate,
\begin{align}
\frac{d^2\Gamma_n}{dE_{p_1}d\cos\theta}=&\frac{G_F^2}{\pi^3}|\bm{p}_1||\bm{p}_2|\sum_{\kappa_1,\kappa_2,\kappa'_1,\kappa'_2,J,l}(2J+1)(2j_{\kappa_1}+1)(2j_{\kappa_2}+1)(2j_{\kappa'_1}+1)(2j_{\kappa'_2}+1)\nonumber\\
&\times\frac{1+(-1)^{l_{\kappa_1}+l_{\kappa'_1}+l}}{2}\frac{1+(-1)^{l_{\kappa_2}+l_{\kappa'_2}+l}}{2}i^{-l_{\kappa_1}-l_{\kappa_2}+l_{\kappa_1'}+l_{\kappa_2'}}e^{i\left(\delta_{\kappa_1}+\delta_{\kappa_2}+\delta_{\kappa_1'}+\delta_{\kappa_2'}\right)} \nonumber\\
&\times(j_{\kappa_1},1/2,j_{\kappa'_1},-1/2|l,0)(j_{\kappa_2},1/2,j_{\kappa'_2},-1/2|l,0)W(j_{\kappa_1}j_{\kappa_2}j_{\kappa'_1}j_{\kappa'_2};Jl) \nonumber\\
&\times(-1)^{J-j_{\kappa_2}-j_{\kappa'_2}}\sum_{i=1}^{6}g_iW_i(J,\kappa_1,\kappa_2,E_{p_1},n)\sum_{i'=1}^{6}g_{i'}^*W_{i'}^*(J,\kappa'_1,\kappa'_2,E_{p_1},n)P_l(\cos\theta),
\end{align}
where $P_l(x)$ is Legendre polynomial and $W$ is Racah coefficient,
and we can estimate the angular integral analytically and obtain the following formula for the decay rate of $\mu^-e^-\rightarrow e^-e^-$ of a muonic atom,
\begin{align}
\Gamma=&\frac{G_F^2}{\pi^3}\sum_{n}\int_{m_e}^{E_\mathrm{max}^n}dE_{p_1}|\bm{p}_1||\bm{p}_2|\sum_{J,\kappa_1,\kappa_2}(2J+1)(2j_{\kappa_1}+1)(2j_{\kappa_2}+1)\left|\sum_{i=1}^{6}g_iW_i(J,\kappa_1,\kappa_2,E_{p_1},n)\right|^2.
\label{eq:result}
\end{align}

\section{Numerical Results \label{sec:Results}}

At first, we study the transition density 
 $\rho_\mathrm{tr}(r)$  given by the product of lepton wave functions as
\begin{eqnarray}
\rho_\mathrm{tr}(r) & = & g_{p_1}^{-1}(r)g_\mu^{-1}(r)g_{p_2}^{-1}(r)g_e^{-1}(r),
\end{eqnarray}
to find  the role of the Coulomb interaction on the lepton wave function.
Here we take the most important transition matrix element of $1S$ electron and muon to 
the $\kappa=-1$ electrons ($\mu^-(1S)+e^-(1S) \rightarrow e^-(\kappa=-1) + e^-(\kappa=-1)$ ),
where the two electrons are equally sharing the energy
 $E_{p_1}=E_{p_2}=(m_\mu+m_e-B^{1S}_\mu-B^{1S}_e)/2$.
We examine four models for the lepton wave functions shown in Table \ref{tab:model}.
In the model I, the bound state wave functions of  muon and electron are
calculated in the non-relativistic approximation with  Coulomb interaction of point nuclear
 charge and the electron scattering states are in the plane wave approximation.
Then the wave function of the scattering state are replaced by the solution of Dirac equation 
in the model II.
In the model III, both of the bound state and the scattering state lepton wave functions are 
calculated from the Dirac equation with point nuclear charge.
Finally, we used the uniform nuclear charge distribution
in the model IV.

\begin{figure}[htb]
\includegraphics[width=70mm]{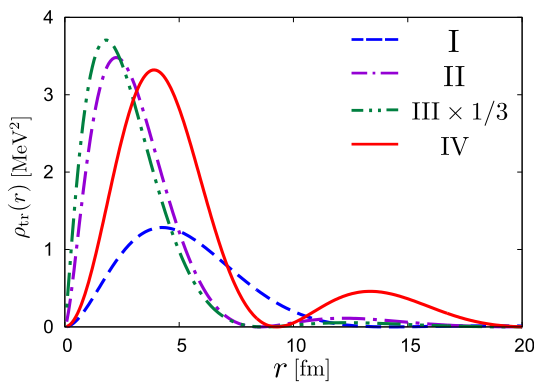}
\caption{The transition density $\rho(r)r^2$ for the $\mu^-(1S)+e^-(1S) \rightarrow e^-(\kappa=-1) + e^-(\kappa=-1)$.
The dashed, dash-dotted, dash-two-dotted and solid curves 
show the transition density in the model I, II, III and IV, respectively.
}
\label{fig:overlap}
\end{figure}

The transition densities of the four models for the $\mu^-e^-\rightarrow e^-e^-$ decay of  $^{208}$Pb muonic atom
are shown in FIG. \ref{fig:overlap}.
The dashed curve shows transition density in the model I which
simulates the previous analysis.
By including the Coulomb attraction for scattering electrons  in the model II,
the transition density is enhanced around the muon Bohr radius as shown by 
the dash-doted curve.
Further we use the consistent lepton wave functions of the Dirac equation 
with point nuclear charge in the model III.
The transition density becomes very large as shown by the dash-two-dotted curve, which is
  $1/3$ of the actual transition density.
However, the use of point nuclear charge would not be appropriate for atom of large $Z$, 
where the Bohr radius of muon can be comparable to the nuclear radius.
The solid curve shows our final result by using finite charge distribution of
nucleus in the model IV.  The peak position  of the
transition density  is shifted toward larger $r$ compared with that of point nuclear charge.
Here the charge distribution of nucleus is taken as uniform distribution as
\begin{align}
\rho_C(r)=\frac{3Ze}{4\pi R^3}\theta(R-r).
\label{eq:uniform}
\end{align}
We use $R=1.2A^{1/3}$ for mass number $A$.
For each $Z$, we take the mass number $A$ of the most abundant isotope \cite{Berglund2011}, 
e.g. $A=208$ for $Z=82$. 
\begin{table}[htb]
\setlength{\extrarowheight}{3pt}
\centering
　\caption{Models for the lepton wave functions.}
　\vspace{0.4cm}
　\doublerulesep 0.8pt \tabcolsep 0.4cm
  \begin{tabular}{ccc} \hline\hline
  Model & Bound state & Scattering state \\ \hline
  I     & Non. Rel./Point Coul. & Rel./PLW \\
  II     & Non. Rel./Point Coul. & Rel./Point Coul. \\
  III     & Rel./Point Coul. & Rel./Point Coul. \\
  IV     & Rel./Uniform Coul. & Rel./Uniform Coul. \\  \hline\hline
  \end{tabular}
  \label{tab:model}
\end{table}

An analytic formula of the $\mu^-e^-\rightarrow e^-e^-$ decay rate of muonic atom %with atomic number $Z$ 
is given in the previous work\cite{Koike2010} as
\begin{align}
\Gamma_0=\frac{m_\mu}{8\pi^2}(Z-1)^3\alpha^3\left(G_Fm_\mu^2\right)^2\left(\frac{m_e}{m_\mu}\right)^3G,
\label{eq:Gamma0}
\end{align}
where $G\equiv G_{12}+16G_{34}+4G_{56}+8G'_{14}+8G'_{23}-8G'_{56}$ 
with $G_{ij}\equiv\left|g_i\right|^2+\left|g_j\right|^2$ 
and $G'_{ij}\equiv\mathrm{Re}\left(g_i^*g_j\right)$.
The formula shows that the decay rate is proportional to $(Z-1)^3$.
The formula was obtained by using the non-relativistic bound state of the muon and $1S$ electron 
with a point nuclear charge and the plane wave approximation for the final electrons.

\begin{figure}[htb]
\includegraphics[width=70mm]{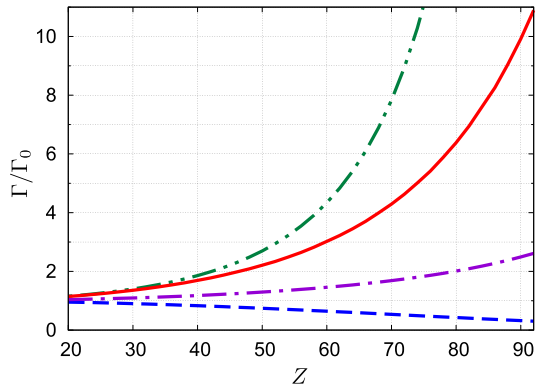}
\caption{The atomic number ($Z$) dependence of the ratio of the decay rate $\Gamma/\Gamma_0$.
The dashed, dash-dotted, dash-two-dotted and solid curves 
show the transition density in the model I, II, III and IV, respectively.
We note that the factor $1/3$ has not been multiplied to the dash-two-dotted curve.}
\label{fig:decayrate}
\end{figure}

The decay rate $\Gamma$ obtained in this work is shown in FIG. \ref{fig:decayrate}.
Here the ratios $\Gamma/\Gamma_0$ are plotted. We retain only the term of $g_1$ and 
set the other $g$'s zero.
The contribution of the dominant $1S$ bound electron is included. 
The dashed  curve in Fig \ref{fig:decayrate}
shows the decay rate evaluated with model I.
The ratio for model I deviates from the unity for large $Z$
because of using finite size bound muon wave function 
instead of using plane wave in the previous estimation.
When we replace the plane wave electrons with the Dirac wave function for point 
nuclear charge(II),  the decay rate increases as shown in the dash-dotted curve.
When both bound and scattering states are described by the Dirac equation that includes
the Coulomb interaction of point nuclear charge(III),
the decay rate is even more enhanced 
as shown in the dash-two-dotted curve.
A realistic description of $\Gamma/\Gamma_0$ is obtained by using the uniform nuclear 
charge distribution in model IV as shown in the solid curve in Fig. \ref{fig:decayrate}.

The results show that, while $\Gamma_0$ gives reasonable estimation for smaller $Z\sim 20$,  
the $Z$ dependence of the $\Gamma$ 
is stronger than $(Z-1)^3$. 
The ratio $\Gamma/\Gamma_0$ is about $7.0$ for the $^{208}$Pb.
We found slightly different $Z$ dependence of $\Gamma$ for two types of the effective CLFV 
contact interaction.
The interaction of $g_i$ term with $i=1,2,3,4$, which leads to the same helicity states of 
two electrons gives  $\Gamma/\Gamma_0 \sim 7.0(1.1)$ for  $^{208}$Pb ($^{40}$Ca).
For $g_i$ term with $i=5,6$, where the opposite helicity states of electrons are emitted, 
the decay rate is $\Gamma/\Gamma_0 \sim 6.3(1.1)$ for  $^{208}$Pb ($^{40}$Ca). 
Therefore $Z$ dependence of the decay rate for  $g_{1} \sim g_{4}$  is 
slightly stronger than that of  $g_{5}$ and $g_{6}$.

All the results of the decay rate were obtained by including sufficiently large number 
of partial waves of final electrons.
The convergence properties of the decay rate against the number of partial waves included
 is shown 
in Table \ref{tab:CaSnPb1}.
The number of partial waves needed to obtain
convergent results  was  $|\kappa| \sim 6$ for Pb and Sn 
and $|\kappa| \sim 13$ for Ca.
This happens because the muon Bohr radius is increasing for decreasing $Z$.
\begin{table}[htb]
\setlength{\extrarowheight}{3pt}
\centering
　\caption{The convergence property of the partial wave expansion of $\Gamma/\Gamma_0$.}
　\vspace{0.4cm}
　\doublerulesep 0.8pt \tabcolsep 0.4cm
  \begin{tabular}{ccccc} \hline\hline
    nuclei & $|\kappa|\le1$ & $|\kappa|\le5$ & $|\kappa|\le10$ & $|\kappa|\le20$ \\ \hline
    $^{40}$Ca & 0.141 & 0.847 & 1.11 & 1.15 \\
    $^{120}$Sn & 0.731 & 2.17 & 2.21 & 2.21 \\
    $^{208}$Pb & 2.89 & 6.94 & 6.96 & 6.96 \\ \hline\hline
  \end{tabular}
  \label{tab:CaSnPb1}
\end{table}

We have also examined realistic form of the distribution of nuclear charge using
the Woods-Saxon form,
\begin{align}
\rho_C(r)=\rho_0\left[1+\exp\left(\frac{r-c}{z}\right)\right]^{-1},
\end{align}
for $^{40}$Ca, $^{120}$Sn and $^{208}$Pb. The parameters, $c$ and $z$ and the ratio of 
the decay rate $\Gamma/\Gamma_0$ are listed in TABLE \ref{tab:CaSnPb2}.
The modification of the decay rate using Woods-Saxon form charge distribution in place of uniform distribution is less than 1\%.
\begin{table}[htb]
\setlength{\extrarowheight}{3pt}
\centering
\caption{The parameters of the charge distribution of Woods-Saxon form and the ratio of 
the decay rates $\Gamma/\Gamma_0$ for $^{40}$Ca, $^{120}$Sn and $^{208}$Pb \cite{Jager1974}.
In the 4th (5th) column, the $\Gamma/\Gamma_0$ including the contribution of the $1S$
($1S$ and higher shells) is shown.}
　\vspace{0.4cm}
　\doublerulesep 0.8pt \tabcolsep 0.4cm
  \begin{tabular}{ccccc} \hline\hline
    nuclei & $c$ [fm] & $z$ [fm] & $\Gamma/\Gamma_0$ (only $1S$) &  $1S+2S+\cdots$ \\ \hline
    $^{40}$Ca & 3.51(7) & 0.563 & 1.15 & 1.35 \\
    $^{120}$Sn & 5.315(25) & 0.576(11) & 2.21 & 2.67 \\
    $^{208}$Pb & 6.624(35) & 0.549(8) & 6.96 & 8.78 \\ \hline\hline
  \end{tabular}
  \label{tab:CaSnPb2}
\end{table}

The results shown so far were obtained including only the main transitions
where the initial electrons are bound in the $1S$ state.
The contributions of the electrons from the higher shell $2S,3S ...$  are estimated within 
the independent particle model for the atomic electrons.
Contributions of higher shell electrons increase the transition rate by $\sim20$\% 
as shown in 
Table \ref{tab:CaSnPb2}, which is consistent with the result of the previous work.

\begin{figure}[htb]
\centering
\includegraphics[width=70mm,clip]{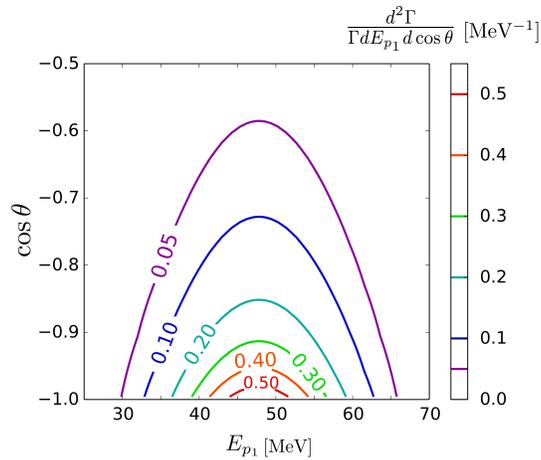}
\caption{The energy and angular distribution of emitted electrons for 
the $^{208}$Pb by using $g_1$ type interaction.}
\label{fig:drdedc}
\end{figure}

\begin{figure}[htb]
\centering
\includegraphics[width=70mm,clip]{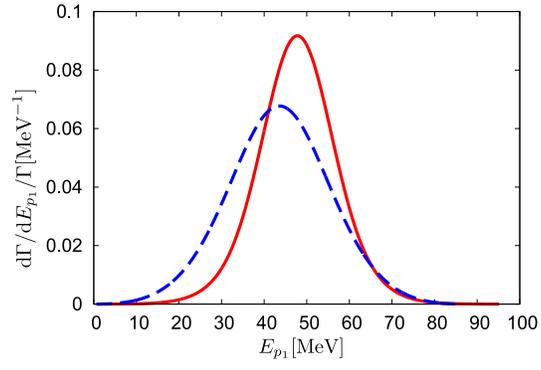}
\caption{The normalized energy distribution of electron for the $^{208}$Pb.
The solid (red) curve and the dashed (blue) curve are
obtained by using model IV and model I, respectively.
The $g_1$ term is included and all bound $S$ state electrons are taken into account.}
\label{fig:drde}
\end{figure}

\begin{figure}[htb]
\centering
\includegraphics[width=70mm,clip]{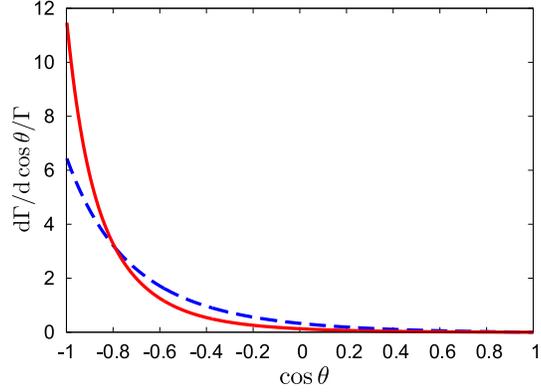}
\caption{The normalized 
angular distribution of emitted electrons for the $^{208}$Pb. 
The other features are the same as those in Fig. \ref{fig:drde}.
}
\label{fig:drdcos}
\end{figure}

%%%%%%%%%%%%%%%%%%%%%%%%%%%%%%%%%%%%%%%%%%%%%%%%%%%%%%%%%%%%%%%%%%%%%%%%
The energy and angular distribution of electron calculated from the 
double differential decay rate  in Eq. (\ref{eq:dgde}) for the $^{208}$Pb is
shown  in FIG. \ref{fig:drdedc}.
The two final electrons are mainly emitted with the same energy in a opposite direction, since
the momentum carried by the bound two leptons is minimized in this configuration.
%%%%%%%%%%%%%%%%%%%%%%%%%%%%%%%%%%%%%%%%%%%%%%%%%%%%%%%%%%%%%%%%%%%
The electron energy spectrum normalized by decay rate $d\Gamma/dE/\Gamma$ 
and the angular distribution between the two electrons 
are shown in FIG. \ref{fig:drde} and  in FIG. \ref{fig:drdcos}
for the models IV(solid) and I(dash).
The maximum of the energy distribution is around half of the total 
energy  $m_\mu+m_e-B^{1S}_\mu-B^{1S}_e$. 
Most of the final electrons are emitted in the opposite directions.
The shapes of the energy distribution 
and the angular distribution are significantly different from the models I and IV.
The angular and the energy distributions in model IV becomes narrower than those of model I.
This is because the muon is less bound for finite range nuclear charge distribution,
and therefore it has smaller high momentum component.

\begin{figure}[htb]
\centering
\includegraphics[width=70mm,clip]{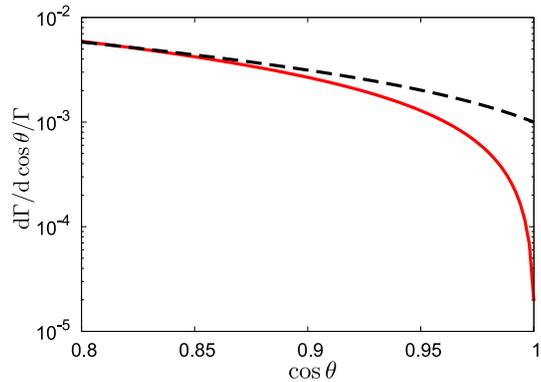}
\caption{The angular distribution of emitted electrons for the $^{208}$Pb.
The angular distribution due to the ($g_1$-$g_4$) terms is shown in solid (red) curve, 
and that of ($g_5$-$g_6$) terms is shown in dashed (black) curve.}
\label{fig:drdcos_log}
\end{figure}

%%%%%%%%%%%%%%%%%%%%%%%%%%%%%%%%%%%%%%%%%%%%%%%%%%%%%%%%%%%%%%%%%%%%%%%
For the interaction which leads to the same chirality of final electrons, 
i.e. $g_1 \sim g_4$ terms of Eq. (4), the Pauli principle prevents the final 
electron from having the same momentum.
On the other hand, in $g_5$ and $g_6$ terms which leads to electrons with opposite chiralities, 
this does not apply.
A difference between two interaction terms appears near $\cos\theta=1$ 
as seen in FIG. \ref{fig:drdcos_log}.

%%%%%%%%%%%%%%%%%%%%%%%%%%%%%%%%%%%%%%%%%%%%%%%%%%%%%%%%%%%%%%%%%%%%%%%%%%%%%%%%%%%%%%%%%%%%%%%%%%

Finally, we evaluate upper limits for the branching ratio of 
the $\mu^-e^-\rightarrow e^-e^-$ decay of a muonic atom.
The branching ratio of $\mu^-e^-\rightarrow e^-e^-$  is defined 
by using the  $\mu^-e^-\rightarrow e^-e^-$ decay rate of muonic atom
 $\Gamma(\mu^-e^-\rightarrow e^-e^-)$ ($\Gamma$ given  in Eq. (6))
and the total decay rate of muonic atom $1/\tilde{\tau}_{\mu}$:
\begin{align}
Br(\mu^-e^-\rightarrow e^-e^-)\equiv \tilde{\tau}_{\mu}\Gamma(\mu^-e^-\rightarrow e^-e^-).
\end{align}
We  estimate the strength of the CLFV interaction 
from the  current upper limit of the branching ratio of  $\mu^+\rightarrow e^+e^+e^-$. 
The branching ratio $Br(\mu^+\rightarrow e^+e^+e^-)$ is given as
\begin{align}
Br(\mu^+\rightarrow e^+e^+e^-)\equiv& \tau_\mu\Gamma(\mu^+\rightarrow e^+e^+e^-).
\end{align}
Here  $\Gamma(\mu^+\rightarrow e^+e^+e^-)$ and $1/\tau_\mu$ are the 
decay rate of  $\mu^+\rightarrow e^+e^+e^-$ 
and total decay rate of free muon, respectively.
Using the contact CLFV interaction in Eq. (\ref{eq:shortrange}), 
the branching ratio $Br(\mu^+\rightarrow e^+e^+e^-)$ is given as\cite{Okada2000}
\begin{align}
Br(\mu^+\rightarrow e^+e^+e^-) =& \frac{1}{8}(G_{12}+16G_{34}+8G_{56}).
\label{eq:br1}
\end{align}
Keeping only $g_{1}$ term of CLFV interaction, we can express the 
branching ratio of $\mu^-e^-\rightarrow e^-e^-$ decay of muonic atom  as\cite{Koike2010} 
\begin{eqnarray}
Br(\mu^-e^-\rightarrow e^-e^-)&\simeq & 192\pi(Z-1)^3\alpha^3\left(\frac{m_e}{m_\mu}\right)^3
\frac{\tilde{\tau}_\mu}{\tau_\mu}\frac{\Gamma}{\Gamma_0} Br(\mu^+\rightarrow e^+e^+e^-).
\end{eqnarray}
Here we used $\tau_\mu=192\pi^3/\left(G_F^2m_\mu^5\right)=2.197\times10^{-6}$ [s].
The upper limit of  $Br(\mu^-e^-\rightarrow e^-e^-)$ can be estimated by
using the current upper limit of the branching ratio $Br(\mu^+\rightarrow e^+e^+e^-)$.

\begin{figure}[htb]
\centering
\includegraphics[width=70mm]{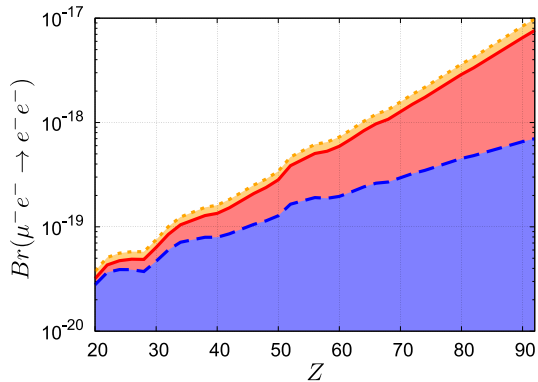}
\caption{Upper limits of $Br(\mu^-e^-\rightarrow e^-e^-)$.
The dashed(blue) curve shows the result of previous work  \cite{Koike2010}.
Our results including only $1S$ electrons and all $1S$ electrons are shown by the solid(red) 
curve and the dotted(orange) curve, respectively.}
\label{fig:branchingratio}
\end{figure}

The upper limits of branching ratio  of the previous work(dashed curve) 
and our results with $1S$(solid curve) and all $nS$ electrons(dotted curve) are 
shown in FIG. \ref{fig:branchingratio}.
Here we used the result of the SINDRUM experiment 
$Br(\mu^+\rightarrow e^+e^+e^-)<1.0\times 10^{-12}$ \cite{Bellgardt1988}
and the data of the lifetime of muonic atoms $\tilde{\tau}_\mu$ given in \cite{Suzuki1987}.
For $^{208}$Pb ($^{238}$U), the branching ratios $Br(\mu^-e^-\rightarrow e^-e^-)$ 
considering only $1S$ electrons 
and all electrons are $3.3\times 10^{-18}$ ($6.9\times 10^{-18}$) and $4.2\times 10^{-18}$ 
($9.8\times 10^{-18}$), respectively. 
$Br(\mu^-e^-\rightarrow e^-e^-)$ reaches about $10^{-17}$ for $^{238}$U.

%It can be suggested that we have the more advantage in the search for the CLFV process as we use the nucleus with the larger proton number.

\section{Conclusion \label{sec:Conclusion}}

% the following is modified by YK 20160226
We have made an improved study on the $\mu^-e^-\rightarrow e^-e^-$ decay in muonic atoms. 
Coulomb interaction of leptons with finite nuclear charge distributions
is taken into account by using the standard multipole expansion formalism 
and the numerical solutions of Dirac equations for both the electron and muon wave functions.
The effects of Coulomb distortion of the emitted electron and relativistic treatments 
of the bound leptons are significantly important
for quantitative estimations of the decay rate.
Enhancements of the decay rates of about 9 and 14 times for  $^{208}$Pb and $^{238}$U respectively compared 
with the previous analysis are obtained due to the enhanced overlap integrals
of the lepton wave functions.
We also found that different helicity structures of the CLFV interaction
generate sizable difference in the $Z$-dependence of the decay rate and also
the angular distribution of the emitted electrons. Finally, the upper limits of the
branching ratio of the $\mu^-e^-\rightarrow e^-e^-$ decay of muonic atom was estimated.

In this work we have included only the four Fermi CLFV interaction.
It is important to estimate the photonic interaction which generates 
long range interactions between the bound muon and many electrons in an atom.
In addition to the decay rates, it would be of great interest to find some other observables that may be useful to discriminate
photonic and contact interactions 
and also various terms of the effective CLFV interactions. These issues are under progress and 
will be discussed in a separate paper.
%the investigation of the model-discriminating 
%power of $\mu^-e^-\rightarrow e^-e^-$ is an interesting topic.
%The candidates for such probes are an energy spectrum of a emitted electron, the angular correlation 
%between two emitted electrons,  helicity, and so on.
%If there are differences in such quantities of several interaction terms, it is expected that the 
%experiment can discriminate the type of the interaction in finding the CLFV process, 
%$\mu^-e^-\rightarrow e^-e^-$ in a muonic atom.

%The remaining important issue is the estimation of the background events.
%For example, the process $\mu^-e^-\rightarrow e^-e^-\nu_\mu\overline{\nu}_e$ can be the candidate.

\begin{acknowledgments}
This work was supported by the JSPS KAKENHI Grant No. 25105009 and 24340044 (J.S.), No. 25105010 (T.S.), No. 25000004 (Y.K.) and No. 25003345 (M.Y.). 
T.S. and Y.U. thank to Dr. S. Nakamura for valuable comments.

\end{acknowledgments}

\end{document}